\documentstyle[12pt,preprint,aps,floats,epsf]{revtex}
\def\partsl{\partial\hspace{-2mm}/}
\def\ppsl{P\hspace{-2.25mm}/\,}
\def\lsl{\ell\hspace{-1.9mm}/\,}

\def\gamsl{\Gamma\hspace{-2.25mm}/\,}
\def\usl{u\hspace{-2.25mm}/\,}
\def\thalf{{\textstyle{\frac{1}{2}}}}
\def\tfith{{\textstyle{\frac{5}{3}}}}
\def\teith{{\textstyle{\frac{8}{3}}}}
\def\toneth{{\textstyle{\frac{1}{3}}}}

\def\tthtwo{{\textstyle{\frac{3}{2}}}}

\newcommand{\be}{\begin{eqnarray}}
\newcommand{\ee}{\end{eqnarray}}
\newcommand{\beq}{\begin{equation}}
\newcommand{\eeq}{\end{equation}}

\begin{document}
\draft
\preprint{NUC-MINN-99/7-T}
\title{Baryon Masses in Chiral Perturbation Theory with Infrared
Regularization}
\author{P. J. Ellis and K. Torikoshi}
\address{School of Physics and Astronomy,
University of Minnesota\\
Minneapolis, MN 55455}
\date{\today}
\maketitle

\begin{abstract}
The baryon masses are examined in $SU(3)$ chiral perturbation 
theory to third order using the recently proposed
infrared regularization scheme. Fourth order is estimated by
evaluating the dominant diagram. With this regularization
the magnitude of the loop integrals 
is reduced so that the convergence of the series appears to 
be better than in the heavy baryon approach. 
\end{abstract}

\newpage

\section{Introduction}

The description of the masses of the baryon octet in $SU(3)$ chiral 
perturbation theory is not in a satisfactory state. Using heavy baryon
chiral perturbation theory (HBChPT) Borasoy and Meissner \cite{bm} have 
carried calculations
to order $Q^4$, where $Q$ denotes a small momentum or a meson mass.
The contributions of order $Q^2$, $Q^3$ and $Q^4$ are of roughly similar
magnitude with alternating sign, thus casting doubt on the convergence of 
the expansion. Donoghue and Holstein \cite{dh} have suggested that
the loop integrals should be regularized by introducing form factors. 
While this can be used to make the loop contributions tiny, it has the 
disadvantage of being model dependent and it is difficult to see how to apply 
it consistently in general situations. 
An alternative approach has recently been suggested 
by Becher and Leutwyler \cite{bl} following the work of Ref. \cite{et}. 
It employs chiral perturbation theory in manifestly Lorentz invariant form
with the loop integrals evaluated in the so-called infrared regularization
(IR) scheme. Our purpose here 
is to examine this scheme in the context of the baryon masses.

In Sec. II we recall the necessary aspects of chiral perturbation theory 
and the infrared regularization scheme. Our results are given in Sec. III, 
first through third order and then including an estimate of fourth order.
Finally our conclusions are presented in Sec. IV.

\section{Theory}

The lowest order $SU(3)$ chiral Lagrangian \cite{kr} is
\begin{equation}
{\cal L}^1={\rm Tr}\left\{i\bar{B}(\partsl B+[\gamsl,B]-M_0
\bar{B}B+\thalf D\bar{B}\gamma_5\{\usl,B\}+\thalf F\bar{B}\gamma_5[\usl,B]
\right\}\;. \label{l1} 
\end{equation}
Here the baryon octet is $B=2^{-\frac{1}{2}}\sum_{a=1}^8\lambda^aB^a$
in terms of the $SU(3)$ matrices $\lambda^a$ and the quantities
involving the meson fields $\phi^a$ are
\begin{eqnarray}
U&=&u^2=\exp\left(\frac{i}{f}\sum_{a=1}^8\lambda^a\phi^a\right)
\nonumber\\
\Gamma_\mu&=&\thalf(u^\dagger\partial_\mu u+u\partial_\mu u^\dagger)
\ ;\ u_\mu=i(u^\dagger\partial_\mu u-u\partial_\mu u^\dagger)\;.
\end{eqnarray}
The second order Lagrangian is 
\begin{equation}
{\cal L}^2=b_0{\rm Tr}(\bar{B}B){\rm Tr}\chi_++b_d{\rm Tr}
(\bar{B}\{\chi_+,B\})+b_f{\rm Tr}(\bar{B}[\chi_+,B])+\ldots\;, 
\label{eq:l2}
\end{equation}
and in $\chi_+=2B_0(u^\dagger{\cal M}u^\dagger+u{\cal M}u)$. In an 
obvious notation the quark mass matrix is
${\cal M}={\rm diag}(\hat{m},\hat{m},m_s)$ and this can be related to the
meson masses to leading order
\begin{equation}
m_\pi^2=2\hat{m}B_0\quad;\quad m_K^2=(\hat{m}+m_s)B_0\quad;\quad
m_\eta^2={\textstyle\frac{1}{3}}(4m_K^2-m_\pi^2)\;.
\end{equation}
For present purposes we can use the above Gell-Mann-Okubo relation 
for the $\eta$ mass; we take $m_\pi=0.139$ GeV and $m_K=0.494$ GeV
yielding $m_\eta=0.565$ GeV.

The order $Q^2$ contribution to the baryon masses take the familiar form
\begin{eqnarray}
M_N(2)&=&-2b_0(2m_K^2+m_\pi^2)-4m_K^2b_d
+4(m_K^2-m_\pi^2)b_f\nonumber\\
M_\Lambda(2)&=&-2b_0(2m_K^2+m_\pi^2)-{\textstyle\frac{4}{3}}
(4m_K^2-m_\pi^2)b_d\nonumber\\
M_\Sigma(2)&=&-2b_0(2m_K^2+m_\pi^2)-4m_\pi^2b_d\nonumber\\
M_\Xi(2)&=&-2b_0(2m_K^2+m_\pi^2)-4m_K^2b_d
+4(m_\pi^2-m_K^2)b_f\;.
\end{eqnarray}

The order $Q^3$ contribution arises from the loop diagram pictured in Fig. 
1 and the guts of this is the integral
\begin{equation}
H=-i\mu^{4-d}\int\frac{d^d\ell}{(2\pi)^d}\frac{1}{(\ell^2-m_M^2+i\epsilon)
(2\ell\cdot P+\ell^2+P^2-M_{B'}^2+i\epsilon)}\;, \label{heq}
\end{equation}
in dimension $d$, with $P$ denoting the four-momentum of the external baryon 
leg, $B$. In \cite{et} it was argued that ``hard" part of the integral,
dominated by poles at momenta of the order of the baryon mass, should be
absorbed in the coefficients of the effective Lagrangian. The ``soft"
part of the integral on the other hand needs to be calculated explicitly
and  can be obtained by expanding out $\ell^2$ from the baryon propagator 
and interchanging the order of integration and summation (the latter 
of course changes the value of the integral).
In dimensional regularization the net effect is that $\ell^2$ is
replaced by $m_M^2$ so that (\ref{heq}) is replaced by 
\begin{equation}
I=-i\mu^{4-d}\int\frac{d^d\ell}{(2\pi)^d}\frac{1}{(\ell^2-m_M^2+i\epsilon)
(2\ell\cdot P+m_M^2+P^2-M_{B'}^2+i\epsilon)}\:. \label{ieq}
\end{equation}
A similar result has been obtained by Becher and Leutwyler \cite{bl}.
They combine the denominators
in Eq. (\ref{heq}) using an integration over a Feynman parameter $z$.
The ``soft" part is then defined by extending the integration range from 
0--1 to 0--$\infty$ and this includes the infrared singular contribution of
leading order $Q^{d-3}$ in the chiral limit $m_M\rightarrow0$ for 
$P^2\simeq M_{B'}^2$. The net 
effect of these maneuvers is to replace the denominator $(ab)^{-1}$
in $H$ by the denominator $[a(b-a)]^{-1}$ to give $I$ in (\ref{ieq}). These 
authors refer to this as infrared regularization. In heavy baryon theory 
the second denominator in (\ref{ieq}) is expanded out thus leaving
$(2\ell\cdot P+i\epsilon)^{-1}$ in leading order. Becher and Leutwyler 
have argued against making this expansion since it breaks down in certain 
regions of parameter space. The present procedure can be viewed as a 
summation of the heavy baryon insertions in the baryon propagator to all 
orders. The leading order term preserves the Weinberg power counting,
but in addition higher order contributions are included.

The integral that needs to be evaluated for the loop diagram of Fig. 1 is 
\begin{equation}
H=-i\mu^{4-d}\int\frac{d^d\ell}{(2\pi)^d}
\frac{\gamma_5\lsl (\ppsl +\lsl +M_{B'})\gamma_5\lsl}{(\ell^2-m_M^2+i\epsilon)
(2\ell\cdot P+\ell^2+P^2-M_{B'}^2+i\epsilon)}\:. \label{hfull}
\end{equation}
We apply the so-called infrared regularization on shell, $P^2=M_B^2$. 
The resulting integrals contain ultraviolet divergences and these are
removed in the standard $\overline{MS}$ scheme; this requires polynomial 
counterterms of arbitrarily high order in $Q$ which we do not need to 
specify. Then Eq. (\ref{hfull}) becomes
\begin{eqnarray}
I(M_B,M_{B'}, m_M)&=&\frac{(M_B^2-M_{B'}^2)m_M^2}{32\pi^2M_B}
\ln\frac{m_M^2}{\mu^2}\nonumber\\
&&+\frac{(M_B+M_{B'})}{16\pi^2M_B}[(M_B-M_{B'})z-m_M^2]J(z,m_M)\;,
\label{eqi}
\end{eqnarray}
where $z=(M_B^2-M_{B'}^2+m_M^2)/(2M_B)$ and
\begin{equation}
J(z,m)=\left\{\matrix{z-z\ln\frac{m^2}{\mu^2}-2\sqrt{m^2-z^2}
\cos^{-1}\left(-\frac{z}{m}\right)&\quad|z|<m\cr
z-z\ln\frac{m^2}{\mu^2}-\sqrt{z^2-m^2}\ln\frac{z+\sqrt{z^2-m^2}}
{z-\sqrt{z^2-m^2}}&\quad|z|>m\;.} \right. \label{jeq}
\end{equation}
This is equivalent to the expression given in Ref. \cite{bl} and is of
leading order $Q^3$. It is natural 
to identify the renormalization scale $\mu$ in this equation with the 
natural scale in the problem, namely the baryon mass in the chiral 
limit, $M_0$. The total third order contribution is then 
$M_B(3)=\sum_{B'M}\alpha_{BB'M}I(M_B,M_{B'}, m_M)$, where the coefficients
$\alpha$ are easily evaluated using Eq. (\ref{l1}) and are listed in the 
Appendix.

To evaluate the baryon masses in fourth order 
${\cal L}^2$ requires additional terms besides those displayed in Eq. 
(\ref{eq:l2}). They contribute to the loop integrals, but 
the value of the coefficients is {\it a priori} unknown. Further,
the finite parts of ${\cal L}^4$ play a role at tree level, again 
introducing unknown coefficients. Borasoy and Meissner \cite{bm}
estimated these low energy constants. They concluded that, while their effect 
is not negligible, the dominant fourth order contribution arises from 
the meson loop diagram pictured in Fig. 2 evaluated using the three terms
of ${\cal L}^2$ given in Eq. (\ref{eq:l2}). Therefore we will consider just 
this contribution in order to get an estimate of the magnitude of fourth 
order. Since only loops containing baryons require special treatment in 
order to keep track of the chiral order, the meson tadpole diagram of Fig. 2 
is evaluated in standard fashion irrespective of whether the IR or HBChPT
schemes are used. For nucleons the result, once divergences have been 
absorbed in the counterterms, is 
\begin{eqnarray}
M_N(4)&=&\biggl\{3(2b_0+b_d+
b_f)m_\pi^4\ln\frac{m_\pi^2}{\mu^2}+
2(4b_0+3b_d-b_f)m_K^4\ln\frac{m_K^2}{\mu^2}+
2b_0m_\eta^4\ln\frac{m_\eta^2}{\mu^2}\nonumber\\
&&\qquad+(-b_d+\tfith b_f)m_\pi^2m_\eta^2\ln\frac{m_\eta^2}{\mu^2}+
\teith(b_d-b_f)m_K^2m_\eta^2\ln\frac{m_\eta^2}{\mu^2}\biggr\}/(4\pi f^2)\;,
\end{eqnarray}
and, as we have remarked, we take $\mu=M_0$.
The contributions for the remaining baryons can be read off from Ref. \cite{bm}.

\section{Results}
\subsection{Through Third Order}

In order to get a first comparison of (\ref{eqi}) with the corresponding
heavy baryon result we simply use the physical masses of the 
particles and pick a reasonable value of the renormalization scale, $\mu=1$ 
GeV. The decay constant $f$ is taken to be the average of the kaon 
and pion decay constants \cite{pdg}, $f=0.103$ GeV. For the parameters $D$ 
and $F$ we take the ratio from Close and Roberts \cite{cr} and their sum
is the axial coupling constant \cite{pdg}, giving $D=0.804$ and $F=0.463$.
The result is labelled IREX in Table 1 and can be compared with the heavy
baryon case, labelled HB, for which Eq. (\ref{eqi}) is simply 
replaced by $m_M^3/(8\pi)$.
The dominant effect comes from the ratio $z/m$ in Eq. (\ref{jeq}) which
is set to zero in the heavy baryon approach, but can be as large as 0.7 
in magnitude here. Clearly it is largest when $B\neq B'$ which is always 
the case
for the kaons, since they carry strangeness, and they dominate numerically.
In comparison to the heavy baryon case the net result is a rather small 
increase for the $N$ and $\Sigma$ and a fairly substantial, $\sim30$\%, 
reduction for the $\Lambda$ and $\Xi$.

Obviously this is not a consistent procedure and since $m_M^2$ is 
of ${\cal O}(Q^2)$ we should treat the baryon masses to the same order.
Thus we should evaluate 
$M_B(3)=\sum_{B'M}\alpha_{BB'M}I(M_0+M_B(2),M_0+M_{B'}(2), m_M)$.
The total baryon mass through third order is then 
$M_B^{\rm tot}=M_0+M_B(2)+M_B(3)$.
Further, as we have remarked, we wish to choose the renormalization scale 
to be the baryon mass in the chiral
limit, {\it i.e.} $\mu=M_0$. In order to disentangle the constants $M_0$ and 
$b_0$ we need a further piece of information for which we select
the $\pi N$ sigma term, 
$\sigma_{\pi N}(0)=\hat{m}\partial M_N^{\rm tot}/\partial\hat{m}$. 
The actual value of the sigma term is not precisely known, but for 
present purposes we will take the currently 
accepted figure of 45 MeV from Gasser {\it et al.} \cite{gls}. This
allows us to determine the unknown parameters $b_0,\:b_d,\:b_f$ 
and $M_0$ by performing 
a least squares fit to the baryon masses and $\sigma_{\pi N}(0)$.
The strange quark contribution to the nucleon mass
$S\equiv m_s\langle N|\bar{s}s|N\rangle=m_s\partial 
M_N^{\rm tot}/\partial\hat{m_s}$
can then be calculated. We obtain $S=360$ MeV. While the 
magnitude of $S$ is poorly known, this appears a little large so we 
constrained $S$ to be 200 MeV in the 
final fit shown under the rubric IR in Tables 1 and 2; this degraded the 
accuracy of the fit to the baryon masses by about 10 MeV. Note that 
a positive value for $S$ is favored here, whereas the
heavy baryon case \cite{bkm} gives a negative value of similar magnitude.
This would suggest larger $KN$ sigma terms here
for which there is some weak experimental support \cite{gen}.
(To obtain the parameters listed in the heavy baryon column of Table 2 the 
masses and $\sigma_{\pi N}(0)$ were
fitted with a resulting $S=-195$ MeV.) 

Table 1 gives some indication of the 
sensitivity to the baryon masses employed. The net loop results evaluated 
in the IR scheme are now in all cases smaller than in the HB case with 
the reduction ranging from 9\% for the $\Sigma$ to 63\% for the $\Lambda$
and the bulk of the effect arises from the 
kaon loop contribution. Note that, referring to Eq. (\ref{jeq}), the 
magnitude of the ratio $z/m$ varies over a wide range, from 0.04 to almost
2. Thus the assumption of the heavy baryon scheme that this ratio is small, 
so that a power series expansion can be made, is questionable.
Table 2 shows that the baryon masses are fit in the IR 
scheme to an accuracy $\sim50$ MeV. More relevant is the ratio of the third 
and second order contributions. In the HB case this ranges from 
1.11 for the $N$ to 0.70 for the $\Sigma$ and $\Xi$, whereas in the IR 
case it ranges from 0.54 for the $\Sigma$ to 0.30 for the $\Lambda$.
The convergence of the expansion is clearly better in the infrared 
regularization scheme, due to the smaller loop integrals.
The values of $\sigma_{\pi N}(0)$
and $S$ used in the fit are uncertain. If $\sigma_{\pi N}(0)$ is 
increased to 55 MeV or $S$ is increased to 300 MeV the fit to the masses
is improved slightly and the ratio of third to second order becomes 
slightly smaller. Conversely for a reduction of $\sigma_{\pi N}(0)$ to
35 MeV or a reduction of $S$ to 100 MeV the results go in the 
opposite direction. The changes are reasonably small so that the tabulated 
case is representative of our results.

\subsection{Fourth Order Estimate}

Here we estimate fourth order by considering the diagram of Fig. 2.\
If this is calculated
using the parameters previously determined, the value of the diagram in 
the IR scheme is 0.17, 0.21, 0.24 and 0.27 GeV for the $N$, 
$\Lambda$, $\Sigma$ and $\Xi$, respectively. For the $N$ and the $\Lambda$
these contributions are similar in magnitude to third order, while for 
the remaining two baryons they are about half of third order. In the HB 
approach the renormalization scale enters for the first time in fourth order 
and, taking $\mu=1$ GeV as in Ref. \cite{bm}, we find 0.26, 0.38, 0.40 
and 0.51 GeV for the $N$, $\Lambda$, $\Sigma$ and $\Xi$, respectively. The 
difference in the magnitudes in the two schemes largely reflects the 
different values chosen for the renormalization scale. 

Since the 
contributions are sizeable, the parameters should be fitted as before
with the fourth order contribution included. The baryon
masses thus become $M_B^{\rm tot}=M_0+M_B(2)+M_B(3)+M_B(4)$.
The results are given in Table 3. In the IR 
approach the fitting produces two minima of similar depth. We reject the 
solution with a rather low mass $M_0$ of 0.463 GeV and a negative value of
$b_d$ and display the solution which appears to evolve
more naturally from the third order parameterization. Here the value of
$M_0$, 0.653 GeV, is about 10\% smaller than in Table 2 which causes
the fourth order contributions to be reduced by about 37\%; the sensitivity
is obviously due to the fact that $M_0$ is becoming comparable to the eta 
and kaon masses. The second and third order contributions display only 
modest changes from Table 2, while the totals give a little better fit to 
the masses here with an average deviation of 40 MeV. In the 
HB case the results displayed correspond to a renormalization scale
$\mu=1$ GeV; they show qualitatively the same trends as the complete fourth 
order calculation of Borasoy and Meissner \cite{bm}. The alternative of
choosing the renormalization scale self-consistently results in 
$\mu=M_0=0.868$ GeV and the fourth order values of Table 3 are decreased
by 17\%, while second order shows a small 5\% increase. Either way the 
strong cancellation, and in some cases overcancellation, between second 
and third order remains. In the IR case however the magnitude of third 
order is about half that of second order, and our fourth order estimate
is smaller on average by roughly a further factor of a half.

It is also interesting to examine the order-by-order contributions to 
the sigma term
\begin{eqnarray}
\sigma_{\pi N}^{IR}&=&0.074-0.036+0.008=0.045\ {\rm GeV}\nonumber\\
\sigma_{\pi N}^{HB}&=&0.059-0.032+0.018=0.045\ {\rm GeV}\;.
\end{eqnarray}
Our values for $\sigma_{\pi N}^{HB}$ are quite close to those reported by 
Borasoy and Meissner \cite{bm}. The convergence of the series for 
$\sigma_{\pi N}$ is similar in the two schemes, with the IR approach weakly 
favored. For completeness we also give breakdown of the strange quark 
contribution to the nucleon mass, while noting that $S$ was fitted in the
IR calculation but left free in the HB case,
\begin{eqnarray}
S^{IR}&=&0.254-0.059+0.005=0.200\ {\rm GeV}\nonumber\\
S^{HB}&=&0.093-0.346+0.230=-0.023\ {\rm GeV}\;.
\end{eqnarray}
While the behavior of the IR series looks much better, we caution that
the very small value of the fourth order contribution is somewhat
fortuitous since it is very sensitive to the value of the scale $\mu$.

\section{Conclusions}

In conclusion we have examined the numerical implications of a new form 
of $SU(3)$ chiral perturbation 
theory where the loop integrals are evaluated using the so-called infrared
regularization scheme. We have examined the chiral series through third 
order and made an estimate of fourth order by evaluating the dominant 
contribution. The most important feature is that the magnitude of the third
order loop integral contribution to the baryon octet masses is smaller
in the infrared scheme than in HBChPT. This means that the strong 
cancellation, in some 
cases overcancellation, of second order that is characteristic of HBChPT
no longer occurs. Thus the convergence of the chiral series appears to be 
better when infrared regularization is used with successive terms 
decreasing in magnitude by about a factor of a half. Given that the 
ratio of the kaon and eta masses to the baryon masses is of this order,
this is probably the most that one could expect.

\section*{Acknowledgements}

We thank the referee for a useful suggestion.
This work  was supported in part by the US Department
of Energy under grant DE-FG02-87ER40328.

\section*{Appendix}

We list here the non-zero coefficients $\alpha$ needed in the 
evaluation of Fig. 1.
\begin{eqnarray}
&&\alpha_{NN\pi}=\alpha_{\Xi\Sigma K}=\tthtwo\alpha_{\Sigma\Xi K}
=-\frac{3(D+F)^2}{4f^2}\nonumber\\
&&\alpha_{N\Sigma  K}=\alpha_{\Xi\Xi\pi}=\tthtwo\alpha_{\Sigma NK}
=-\frac{3(D-F)^2}{4f^2}\nonumber\\
&&\alpha_{NN\eta}=\alpha_{\Xi\Lambda K}=\thalf\alpha_{\Lambda\Xi K}=
-\frac{(D-3F)^2}{12f^2}\nonumber\\
&&\alpha_{N\Lambda K}=\alpha_{\Xi\Xi\eta}=\thalf\alpha_{\Lambda NK}
=-\frac{(D+3F)^2}{12f^2}\nonumber\\
&&\alpha_{\Lambda\Lambda\eta}=\alpha_{\Sigma\Sigma\eta}=
\alpha_{\Sigma\Lambda\pi}=\toneth\alpha_{\Lambda\Sigma\pi}=
-\frac{D^2}{3f^2}\nonumber\\
&&\alpha_{\Sigma\Sigma\pi}=-\frac{2F^2}{f^2}\;.
\end{eqnarray}

\newpage
\begin{center}
{\bf Table 1}\hfill\\
Meson contributions to the loop result for the baryon masses
(in GeV).
\end{center}
\begin{center}
\begin{tabular}{|l|l|rrr|r|} \hline
&Case& $\pi$& $K$ & $\eta$ &Total\\ \hline
$N$&HB&$-0.012$&$-0.221$&$-0.019$&$-0.252$\\
&IREX&$-0.011$&$-0.249$&$-0.018$&$-0.278$\\
&IR&$-0.012$&$-0.129$&$-0.019$&$-0.160$\\ \hline
$\Lambda$&HB&$-0.007$&$-0.388$&$-0.146$&$-0.540$\\
&IREX&$-0.009$&$-0.195$&$-0.139$&$-0.342$\\
&IR&0.015&$-0.069$&$-0.147$&$-0.201$\\ \hline
$\Sigma$&HB&$-0.006$&$-0.389$&$-0.146$&$-0.541$\\
&IREX&$-0.004$&$-0.406$&$-0.140$&$-0.550$\\
&IR&$-0.008$&$-0.339$&$-0.147$&$-0.494$\\ \hline
$\Xi$&HB&$-0.001$&$-0.557$&$-0.271$&$-0.829$\\
&IREX&$-0.001$&$-0.359$&$-0.262$&$-0.622$\\
&IR&$-0.001$&$-0.314$&$-0.274$&$-0.589$\\ \hline
\end{tabular}
\end{center}
\newpage
\begin{center}
{\bf Table 2}\hfill\\
Comparison of parameters and baryon masses obtained in the heavy
baryon and the present approach to third order.
\end{center}
\begin{center}
\begin{tabular}{|ll|rr|} \hline
\multicolumn{2}{|l|}{Parameters}&HB&IR\\ \hline
\multicolumn{2}{|l|}{$b_0\ ({\rm GeV}^{-1})$}&$-0.762$&$-0.921$\\
\multicolumn{2}{|l|}{$b_d\ ({\rm GeV}^{-1})$}&0.067&0.206\\
\multicolumn{2}{|l|}{$b_f\ ({\rm GeV}^{-1})$}&$-0.533$&$-0.435$\\ 
\multicolumn{2}{|l|}{$M_0 \ ({\rm GeV})$}&0.965&0.733\\ \hline
\multicolumn{2}{|l|}{Masses (GeV)}&&\\ \hline
&second&0.228&0.342\\
$N$&third&$-0.252$&$-0.160$\\
   &total&0.941&0.915\\ \hline
&second&0.687&0.671\\
$\Lambda$&third&$-0.540$&$-0.201$\\
   &total&1.112&1.204\\ \hline
&second&0.768&0.919\\
$\Sigma$&third&$-0.541$&$-0.494$\\
   &total&1.191&1.158\\ \hline
&second&1.187&1.124\\
$\Xi$&third&$-0.829$&$-0.589$\\
   &total&1.322&1.268\\ \hline
\end{tabular}
\end{center}
\newpage
\begin{center}
{\bf Table 3}\hfill\\
Comparison of parameters and baryon masses obtained in the heavy
baryon and the present approach to fourth order.
\end{center}
\begin{center}
\begin{tabular}{|ll|rr|} \hline
\multicolumn{2}{|l|}{Parameters}&HB&IR\\ \hline
\multicolumn{2}{|l|}{$b_0\ ({\rm GeV}^{-1})$}&$-0.577$&$-0.830$\\
\multicolumn{2}{|l|}{$b_d\ ({\rm GeV}^{-1})$}&0.054&0.159\\
\multicolumn{2}{|l|}{$b_f\ ({\rm GeV}^{-1})$}&$-0.424$&$-0.401$\\ 
\multicolumn{2}{|l|}{$M_0 \ ({\rm GeV})$}&0.849&0.653\\ \hline
\multicolumn{2}{|l|}{Masses (GeV)}&&\\ \hline
&second&0.152&0.327\\
$N$&third&$-0.252$&$-0.168$\\
   &fourth&0.192&0.111  \\
   &total&0.941&0.924\\ \hline
&second&0.517&0.640\\
$\Lambda$&third&$-0.540$&$-0.246$\\
 &fourth&0.287&0.136  \\
   &total&1.112&1.184\\ \hline
&second&0.581&0.830\\
$\Sigma$&third&$-0.541$&$-0.476$\\
&fourth&0.303&0.152  \\
   &total&1.191&1.160\\ \hline
&second&0.914&1.048\\
$\Xi$&third&$-0.829$&$-0.591$\\
&fourth&0.389&0.169  \\
   &total&1.322&1.280\\ \hline
\end{tabular}
\end{center}
\newpage
\begin{figure}
\setlength\epsfxsize{7.5cm}
\centerline{\epsfbox{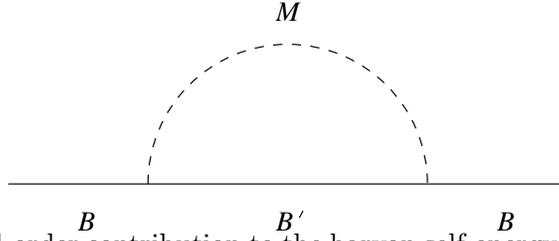}}
\caption{One-loop third-order contribution to the baryon self-energy. The 
dashed line 
represents a meson propagator and the baryons are denoted by solid lines.}
\end{figure}
\begin{figure}
\setlength\epsfxsize{7.5cm}
\centerline{\epsfbox{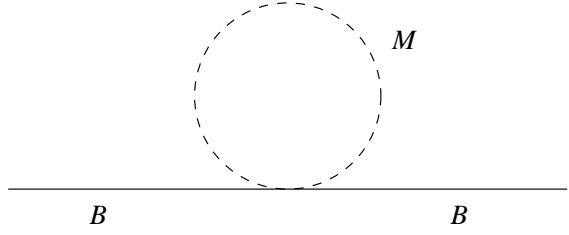}}
\caption{One-loop fourth-order contribution to the baryon self-energy;
notation as for Fig. 1.}
\end{figure}
\centering \leavevmode
\end{document}